\documentclass[]{spie}
\usepackage[utf8]{inputenc}

\usepackage{amsmath,amsfonts,amssymb}
\usepackage{graphicx}
\usepackage[colorlinks=true, allcolors=blue]{hyperref}
\usepackage{aas_macros}

\usepackage[normalem]{ulem}
\newcommand{\tcm}{21$\,$cm~}  %
\newcommand{\HI}{\ensuremath{{\rm HI}}}

\begin{document}

\title{Characterization of the John A. Galt telescope for radio holography with CHIME}

\newcommand{\UBC}{Department of Physics and Astronomy, University of British Columbia, Vancouver, BC, Canada}
\newcommand{\MITP} {Department of Physics, Massachusetts Institute of Technology, Cambridge, MA, USA}
\newcommand{\MITK} {MIT Kavli Institute for Astrophysics and Space Research, Massachusetts Institute of Technology, Cambridge, MA, USA}
\newcommand{\TRU}{Department of Physical Sciences, Thompson Rivers University, Kamloops, BC, Canada}
\newcommand{\PI}{Perimeter Institute for Theoretical Physics, Waterloo, ON, Canada}
\newcommand{\DRAO}{Dominion Radio Astrophysical Observatory, Herzberg Astronomy \& Astrophysics Research Centre, National Research Council Canada, Penticton, BC, Canada}
\newcommand{\UBCO}{Department of Computer Science, Math, Physics, and Statistics, University of British Columbia-Okanagan, Kelowna, BC, Canada}
\newcommand{\McGill}{Department of Physics, McGill University, Montreal, QC, Canada}
\newcommand{\UofTastro}{David A.\ Dunlap Department of Astronomy \& Astrophysics, University of Toronto, Toronto, ON, Canada}
\newcommand{\UofTphys}{Department of Physics, University of Toronto, Toronto, ON, Canada}
\newcommand{\WVU} {Department of Computer Science and Electrical Engineering, West Virginia University, Morgantown WV, USA}
\newcommand{\WVUA} {Department of Physics and Astronomy, West Virginia University, Morgantown, WV, USA}
\newcommand{\WVUGWAC} {Center for Gravitational Waves and Cosmology, West Virginia University, Morgantown, WV, USA}
\newcommand{\Yale}{Department of Physics, Yale University, New Haven, CT, USA}
\newcommand{\YaleA}{Department of Astronomy, Yale University, New Haven, CT, USA}
\newcommand{\Dunlap}{Dunlap Institute for Astronomy and Astrophysics, University of Toronto, Toronto, ON, Canada}
\newcommand{\RRI}{Raman Research Institute, Sadashivanagar,   Bengaluru, India}
\newcommand{\ASIAA}{Institute of Astronomy and Astrophysics, Academia Sinica, Taipei, Taiwan}
\newcommand{\CITA}{Canadian Institute for Theoretical Astrophysics, Toronto, ON, Canada}
\newcommand{\CIFAR}{Canadian Institute for Advanced Research, 180 Dundas St West, Toronto, ON, Canada }
\newcommand{\WVUphysastro} {Department of Physics and Astronomy, West Virginia University, Morgantown, WV, USA}
\newcommand{\Kapteyn}{Kapteyn Astronomical Institute, University of Groningen, Groningen, The Netherlands}

\author[a]{Alex Reda}
\affil[a]{\Yale}
\author[b]{Tristan Pinsonneault-Marotte}
\affil[b]{\UBC}
\author[c,d]{Meiling Deng}
\affil[c]{\DRAO}
\affil[d]{\PI}
\author[b]{Mandana Amiri}
\author[e,f]{Kevin Bandura}
\affil[e]{\WVU}
\affil[f]{\WVUGWAC}
\author[g]{Arnab Chakraborty}
\affil[g]{\McGill}

\author[c,d]{Simon Foreman}
\author[b]{Mark Halpern}
\author[c,h]{Alex S. Hill}
\affil[h]{\UBCO}
\author[b,i]{Carolin H\"ofer}
\affil[i]{\Kapteyn}
\author[j]{Joseph Kania}
\affil[j]{\WVUphysastro}
\author[c]{T.L. Landecker}
\author[b]{Joshua MacEachern}
\author[k,l]{Kiyoshi Masui}
\affil[k]{\MITK}
\affil[l]{\MITP}
\author[j]{Juan Mena-Parra}
\author[b]{Nikola Milutinovic}
\author[a]{Laura Newburgh}
\author[c,h]{Anna Ordog}
\author[g]{Sourabh Paul}

\author[b]{J. Richard Shaw}
\author[g]{Seth R. Siegel}
\author[b]{Rick Smegal}
\author[k,l]{Haochen Wang}
\author[g]{Dallas Wulf}

\authorinfo{Corresponding author: Alex Reda, alex.reda@yale.edu}

\maketitle

\begin{abstract}
    The Canadian Hydrogen Intensity Mapping Experiment (CHIME) will measure the \tcm 
    emission of astrophysical neutral hydrogen to probe large scale structure at redshifts $z=0.8$--2.5. However, detecting the \tcm signal beneath substantially brighter foregrounds remains a key challenge. Due to the high dynamic range between \tcm and foreground emission, an exquisite calibration of instrument systematics, notably the telescope beam, is required to successfully filter out the foregrounds. One technique being used to achieve a high fidelity measurement of the CHIME beam is radio holography, wherein signals from each of CHIME’s %
    analog inputs are correlated with the signal from a co-located reference antenna, the 
    26\,m John A. Galt telescope, as the 26\,m Galt telescope %
    tracks a bright point source transiting over CHIME. %
    In this work we present an analysis of several of the Galt telescope's properties.  We employ drift-scan measurements of several bright sources, along with background estimates derived from the 408\,MHz %
    Haslam map, to estimate the Galt system temperature. To determine the Galt telescope's beam shape, we perform and analyze a raster scan of the bright radio source Cassiopeia A. Finally, we use early holographic measurements to measure the Galt telescope's geometry with respect to CHIME for the holographic analysis of the CHIME-Galt interferometric data set.  
\end{abstract}

\section{Introduction}

The Canadian Hydrogen Intensity Mapping Experiment (CHIME) is a transit interferometer located at the Dominion Radio Astrophysical Observatory (DRAO) near Penticton, British Columbia, Canada. The instrument consists of four 20\,m wide by 100\,m long, stationary, parabolic cylindrical dishes, each outfitted with 256 dual-polarization cloverleaf antennas. Additional details about the specifications of the instrument can be found in \cite{chimeoverview}.

CHIME, which observes between 400--800\,MHz, is designed to measure the \tcm line emission of neutral hydrogen (\HI) from galaxies at redshifts $z=0.8$--2.5. Intensity mapping of the \tcm line 
represents an important developing frontier of cosmology. As a tracer of the underlying matter distribution across redshifts, the \tcm intensity field contains rich information about the Universe's large scale structure and its time evolution \cite{petersontcm, astro-ph/0312134, highztcmroadmap, liushawtcmbible}. 
Numerous experiments, including CHIME, are already underway, or on the horizon, pursuing this signal to shed light on pressing cosmic mysteries (see e.g. \cite{LEDAintro, LOFARintro, HERAintro, HIRAXintro, chimedetection, GBTresult, GBT2022ebossxcorr, GMRTresult, meerkatresult, paperresult, edgesresult, saras3result}).

Specifically, CHIME is pursuing the \tcm signal as a probe of the baryon acoustic oscillations (BAOs), a signature of primordial structure formation imprinted on the large scale distribution of galaxies today and observable in the matter (and hence, 21\,cm) power spectrum. BAOs are of interest due to their utility as a standard ruler, a scale of known comoving size whose evolution with cosmic time can be used to trace the expansion history of the Universe \cite{Seo:2003, Seo:2007}. 

It was first noted in a paper by Riess et al. \cite{1998:Riess} that the expansion of the Universe is accelerating. This accelerated expansion is attributed to ``dark energy"   which is estimated to make up on the order of 70\% of the energy density of the Universe, yet is not well understood by fundamental physics due to a paucity of constraints on the expansion history within the so-called ``redshift desert" between $z \sim 1$--3, when dark energy was becoming the dominant component of the Universe. 

Traditional galaxy surveys targeting BAO have limited statistical sensitivity due to the necessity of optically confirming individual objects. On the other hand, in \tcm intensity mapping, one integrates the emission over many galaxies in lower angular resolution surveys, thereby targeting directly the distribution of neutral hydrogen as a (biased) tracer of the matter distribution on the large cosmological scales most relevant for study of BAOs. By carrying out this measurement of the BAO standard ruler over the critical redshift range $z=0.8$--2.5, CHIME will provide state-of-the-art constraints on models of dark energy and the accelerated expansion of the Universe.

One of the common obstacles facing all \tcm intensity mapping experiments, including CHIME, is that the cosmological \tcm signal is as many as 5 orders of magnitude dimmer than the astrophysical foreground signals, dominated by Galactic and extragalactic synchrotron emission. %
These foregrounds are spectrally distinct from the \tcm signal; while the foregrounds are smooth in frequency, the \tcm signal, which probes different structures along the line of sight across redshifts, decorrelates in frequency. A number of foreground filtering and avoidance schemes have appeared in the literature exploiting these properties \cite{foregrounddayenu, liushawtcmbible, foregroundgprfilter, foregroundkpca} %
, but these methods are easily compromised by an incomplete understanding of instrument systematics, including the angular response (the beam of a radio telescope). As the beam is itself frequency dependent, often times in non-trivial ways, the foregrounds as observed by the instrument are no longer spectrally smooth and naive foreground removal schemes fail. 
To remedy this, instrumental systematics must be modeled to fairly high precision; an early study of CHIME found that if the only source of instrumental uncertainty were the per-feed beam-widths, those widths would have to be constrained to within 0.1\% to enable an unbiased measurement of the \tcm power spectrum \cite{Shawmmodes}. %

\subsection{The holographic beam mapping technique}

One of the methods being used to perform this beam calibration is radio holography. Radio holography is a well-established technique for the metrology of large reflector (radio wave) antennas, which due to their size are typically constructed by combining many individually fabricated reflective panels. In its typical application, holography is used for inferring imperfections and errors in the alignment of these substructures by exploiting the Fourier transform relationship between the antenna farfield pattern and the aperture field; assuming a perfectly constructed reflector, the phase of the complex aperture field should be uniform. Deviations from uniformity are then understood as originating from surface alignment errors that can then be corrected \cite{holographyscottryle, holography30m, holographyeffelsberg100m, holographyyebes40m}. 

The possibility of recovering this information was raised by Silver in 1949 in the text \textit{Microwave Antenna Theory and
Design}, with the caveat that uniquely determining the aperture phase from the Fourier transform requires a measurement of the farfield phase, unobtainable in single dish measurements. In a 1966 paper by P. Smith \cite{1966holotheory}, the theory of using an interferometer to obtain the needed phase information was first presented, and the first successful applications of the technique followed in the 1970s (see the review by Baars\cite{baarsholohistory} and references therein).

The application of the holographic technique to cylindrical dishes was first demonstrated on the CHIME Pathfinder in 2016 by Berger et al. \cite{bergerholo}. For beam calibration for CHIME, holography is performed by cross-correlating the signals measured by each CHIME feed with a ``reference" signal provided by the co-located John A. Galt (hereafter, Galt or 26\,m) telescope, an equatorially mounted 26\,m diameter dish which tracks a chosen celestial calibrator source transiting over DRAO. This correlation traces the 
far field beam response for 
each CHIME feed illuminating the dish individually, at all frequencies and polarizations, along a track at the declination of the calibrator source. Observations of multiple calibrator sources at different declinations provide a sparse sampling of the full 2D beam shape in the declination direction and dense sampling along the hour angle direction. 
Despite offering relatively limited North-South information, holography offers several unique benefits to the CHIME beam-mapping effort. Holography provides full phase information, which can be used to understand the physical features of the reflector as in traditional holography; by providing the beam response of all feeds, holography can be used to assess non-redundancies in the CHIME array; because the Galt telescope's %
field of view is much smaller than that of CHIME, high signal-to-noise measurements of the beam response can be obtained in the far sidelobes by circumventing confusion noise due to CHIME's large field of view. 

A presentation of the first results of CHIME's holographic beam-mapping effort is currently in preparation. In support of that work, in what follows in these proceedings we present an analysis of the 
Galt telescope used as the reference for CHIME holography. 

\section{The John A. Galt 26m Telescope}
The John A. Galt 26\,m telescope is an equatorially-mounted paraboloidal reflector antenna located at DRAO \cite{tomgalt}. The reflector has $f / D = 0.3$, i.e. the focal length $f = 7.6$\,m. The Galt telescope, built in 1959, has been outfitted with a variety of instruments for various measurements. For a recent example see the Global Magneto-Ionic Medium Survey (GMIMS) project, a wideband survey of the Faraday depth of the Galactic interstellar medium using the Galt telescope \cite{GaltGMIMS}. 

When the Galt telescope is being used for holography of CHIME, it is outfitted with CHIME-specific receiving hardware, including the dual polarization cloverleaf antenna \cite{meilingclover} and CHIME-customized filters and amplifiers. The cloverleaf antenna is the same as those used in full CHIME, and includes an additional port allowing injection of a calibration signal from a thermally regulated noise diode.
This signal is not used in the following analysis. To circularize the feed beam pattern for illuminating the axially symmetric Galt reflector, and limiting spillover with an 11\,dB edge taper, a sleeve and choke are attached to the cloverleaf. See Figure \ref{fig:feedsleeve} for the schematic of the sleeve. The feed structure is then encased in a radio-transparent polyethylene cap for waterproofing. The feed is aligned such that when observing a source at the zenith of CHIME its two dipole axes are aligned with those of CHIME; in what follows ``Y polarization" will refer to the North-South oriented dipole and ``X polarization" to the East-West oriented dipole. 

\begin{figure*}
    \centering
    \includegraphics[width=1.0\linewidth]{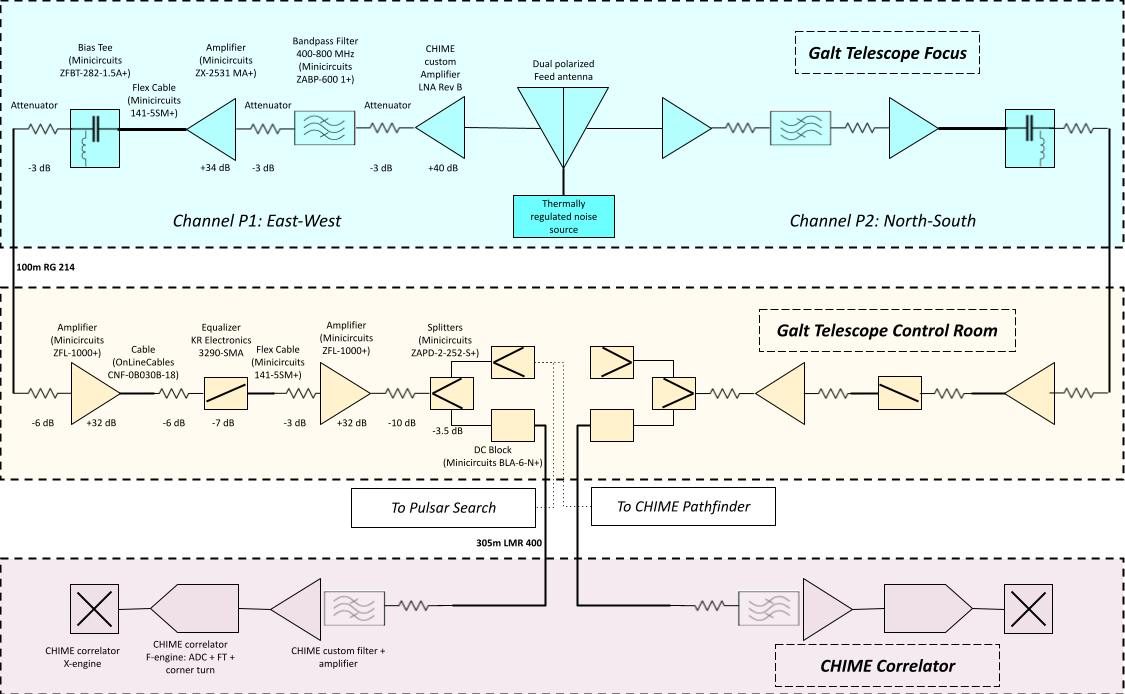}
    \caption{The signal chain of the Galt telescope, as configured for holography with CHIME. The receiver can at the focus of the Galt telescope houses the feed and custom CHIME electronics including the LNA and FLA. This is followed by a chain of commercial electronics and cables, which culminates in one of several science back-ends.} 
    \label{fig:26signalchain}
\end{figure*}

Following the feed is a chain of amplifiers (see Figure \ref{fig:26signalchain}). First is a proprietary low noise amplifier (LNA) designed and built by CHIME 
(noise temperature $\sim 20-25$ K), followed by a proprietary filter which restricts the signal to the 400--800 MHz 
band used by CHIME. The signal is amplified once more by a commercial amplifier before being sent over 100\,m of coaxial cable to the Galt control room where it is amplified again and equalized (to account for cable losses) and split to be delivered to several downstream backends, including the CHIME correlator following 305\,m of coaxial cable. 

At the FX correlator \cite{CHIMEICE, CHIMEGPU}, the signal undergoes the same digitization and channelization as all other CHIME signals prior to being correlated with those signals to form the holographic visibilities. These will be discussed in Section \ref{section:galtgeo}; in this work we will mostly be concerned with the autocorrelations of the Galt signals.

At this stage the 400 MHz bandwidth is divided into 1024 channels of equal bandwidth $\Delta\nu \sim 390$ kHz. All holographic data products, including the Galt autocorrelations used in the following analysis, are integrated to a $\sim 5$ second cadence in the correlator before being written to disk and served to long-term offline storage on The Digital Research Alliance of Canada's Cedar cluster \cite{chimeoverview}. 

\begin{figure}
    \centering
    \includegraphics[width=1.0\linewidth]{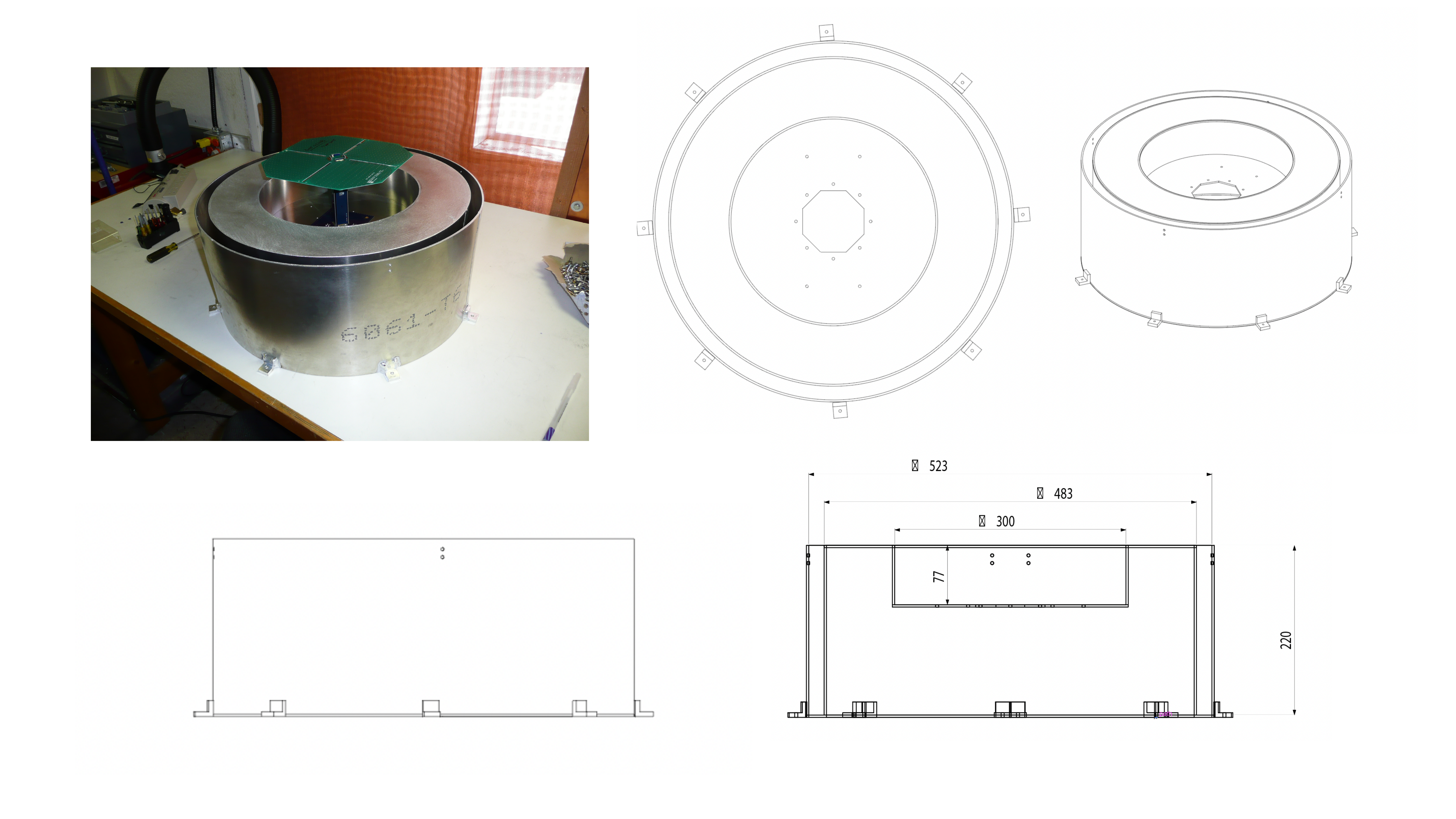}
    \caption{The CHIME cloverleaf feed within the sleeve structure (top left), illustrated schematically. The CHIME feed was designed to have a very wide beam pattern to fully illuminate the $f / D = 0.25$ CHIME cylindrical reflectors. The sleeve is used to circularize and taper the feed pattern, making it suitable for use on the Galt reflector. All dimensions are specified in units of millimeters.}
    \label{fig:feedsleeve}
\end{figure}

\section{Galt beam}

\label{section:galtbeam}

To measure the beam shape of the Galt telescope, we performed a raster scan of the bright radio calibrator source Cassiopeia A (Cas A). Cas A is the second brightest radio source observable from DRAO within CHIME's band,
about 2400 Jy at 600 MHz, 
dimmer only than Cygnus A. This makes it an ideal target for beam calibration; we confirmed that there is no compression of the autocorrelation channels while observing Cas A \cite{juanquantizationbias}, which is not true of Cygnus A. 

We expect the FWHM of the Galt telescope's beam to be 1--2$^\circ$ in the 400--800 MHz band. We therefore performed the scan in two phases on a $8^\circ$x$8^\circ$ grid centered on Cas A, with the Galt telescope first scanning continuously in declination (at fixed right ascension) %
and then in right ascension (at fixed declination). %
In each phase the telescope swept across the source at a rate of 0.5 arcminutes per second while continuously taking data 
in slices of constant declination (right ascension). The grid spacing of the slices is 30 arcminutes in each scanning direction. Radio receiver autocorrelation data is collected at roughly 5 second cadence while pointing data (taken from a different source) is read-in at exactly 5 second cadence. %

\begin{figure*}
    \centering
    \includegraphics[width=1.0\linewidth]{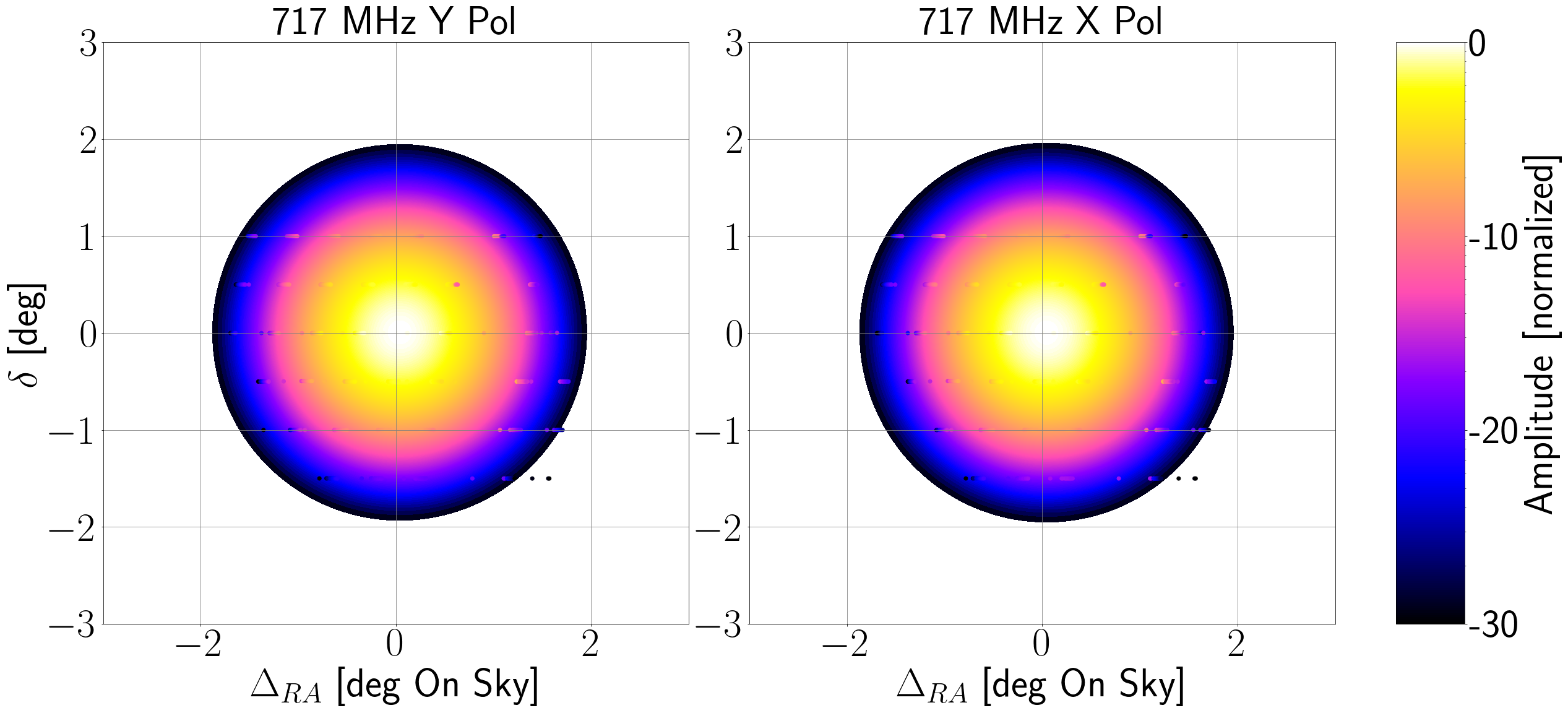}
    \caption{The Galt beam response at 717\,MHz, in X and Y polarization, centered on the location of Cas A (the right ascension axis has been corrected by the cosine of the declination). The contours are evaluated from the best-fit model to the data, overplotted as colored dots. The sky background has been removed and the best fit amplitude has been divided out for normalization. At this frequency the beams are strikingly circular, with FWHM $\sim 1.3^\circ$.}
    \label{fig:galtbeam717}
\end{figure*}

We present here an analysis of a limited subset of the full raster scan dataset. Modeling the Galt beam from this data is complicated by the slewing motions of the raster scan, i.e. when the telescope has completed scanning along a particular declination (right ascension) slice and rapidly moves back across the source to begin scanning along the next slice. Here the telescope moves quickly across the sky relative to the 5 second radio receiver acquisition, so we locate and manually flag out these intervals in the data prior to further analysis. 
As this behavior is more easily tracked in the right ascension phase, we omit the declination phase of the scan in this analysis. 
This leaves us with a square grid covering $4^\circ$ on each side around the source (see Figure \ref{fig:galtbeam717}), which is well sampled in the RA direction and more sparsely sampled in the declination direction. However, this is adequate for our purposes, as we are interested only in a best-fit model of the Galt beam, for which this sampling is still sufficient, rather than a full-grid high resolution measurement. 

\begin{figure*}
    \centering
    \includegraphics[width=1.0\linewidth]{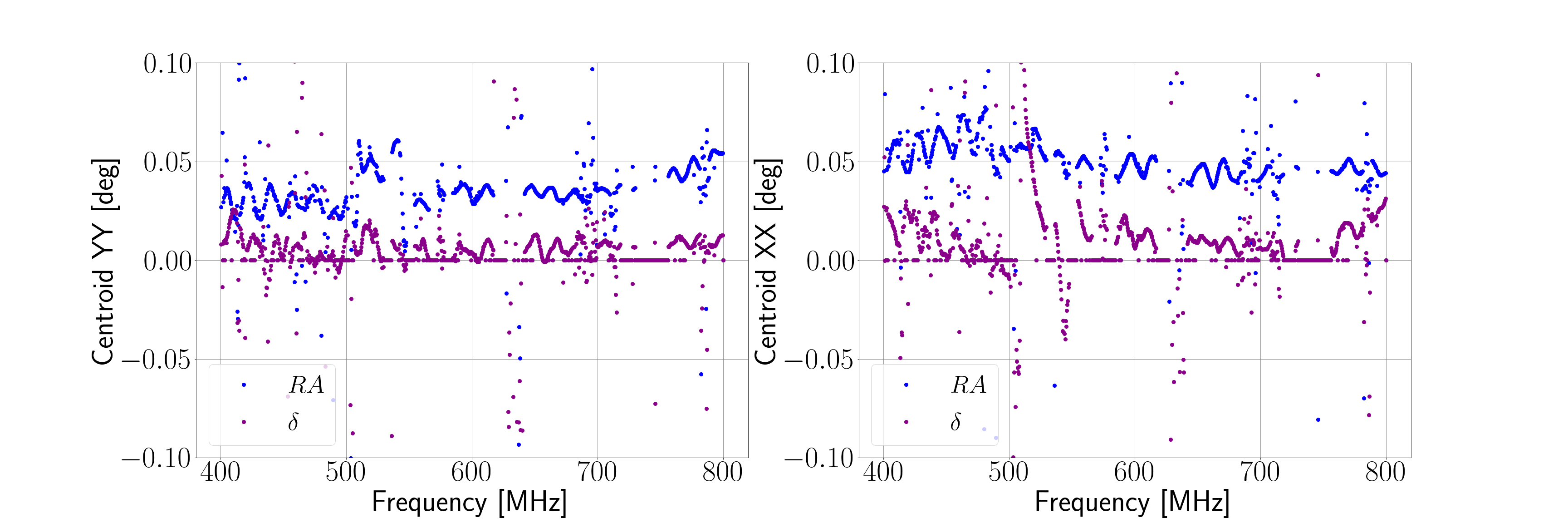}
    \caption{The best-fit centroids of the Galt beam response. The pointing is systematically offset from boresight by less than $.05^\circ$. The variation across frequency is small, $\lesssim .03^\circ$ rms, and is dominated by a spectral ripple of $\sim$ 19\,MHz periodicity.}
    \label{fig:centroids}
\end{figure*}

To model the main 
beam, based on CST simulations of the compound feed-sleeve structure, 
we choose a two-dimensional Gaussian profile. We include terms allowing for both a flat and spatially (linearly) varying background behind the source to account for the location of Cas A on the Galactic plane. We fit our selected data to the coordinates recorded by the Galt telescope's pointing readout system. These coordinates are first corrected given a pointing model for the Galt telescope. In addition, we find that there is an offset between the timestamps recorded by the Galt pointing monitor and those of the raw data stream. To account for this, when fitting we interpolate the pointing onto the timestamps provided with the radio receiver autocorrelation 
data. This interpolation includes an overall time offset which is implemented as an additional free parameter in the model and marginalized over (in practice these offsets were found to be $\sim40$ seconds). We perform the fit at all frequencies and both polarization channels independently. We note here that the Galt pointing, inferred through the best fit centroids, is slightly offset from boresight by less than $.05^\circ$. The centroid variation across frequency is also small, $\lesssim .03^\circ$ rms. This variation is dominated by a sinusoidal ripple with periodicity of about 19\,MHz. The recovered centroids are shown in Figure \ref{fig:centroids}.

\begin{figure*}
    \centering
    \includegraphics[width=1.0\linewidth]{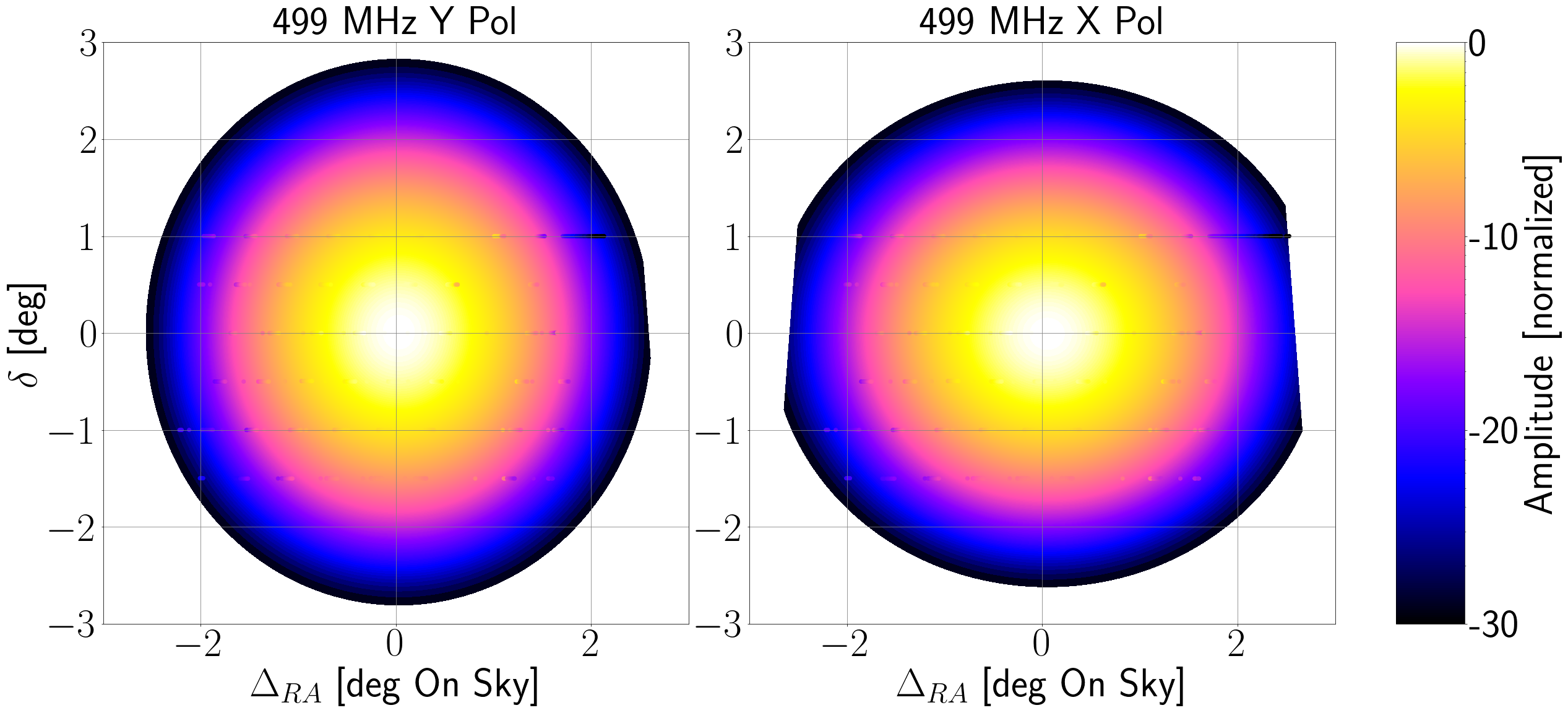}
    \caption{The Galt beam response at 499\,MHz, in X and Y polarization. At this lower frequency, the beams are expectedly wider due to diffraction, but are also noticeably elliptical. This frequency was selected to be representative of the most dramatic excursions from circularity seen in the best-fit model.}
    \label{fig:galtbeam499}
\end{figure*}

\begin{figure*}
    \centering
 \includegraphics[width=.95\linewidth]{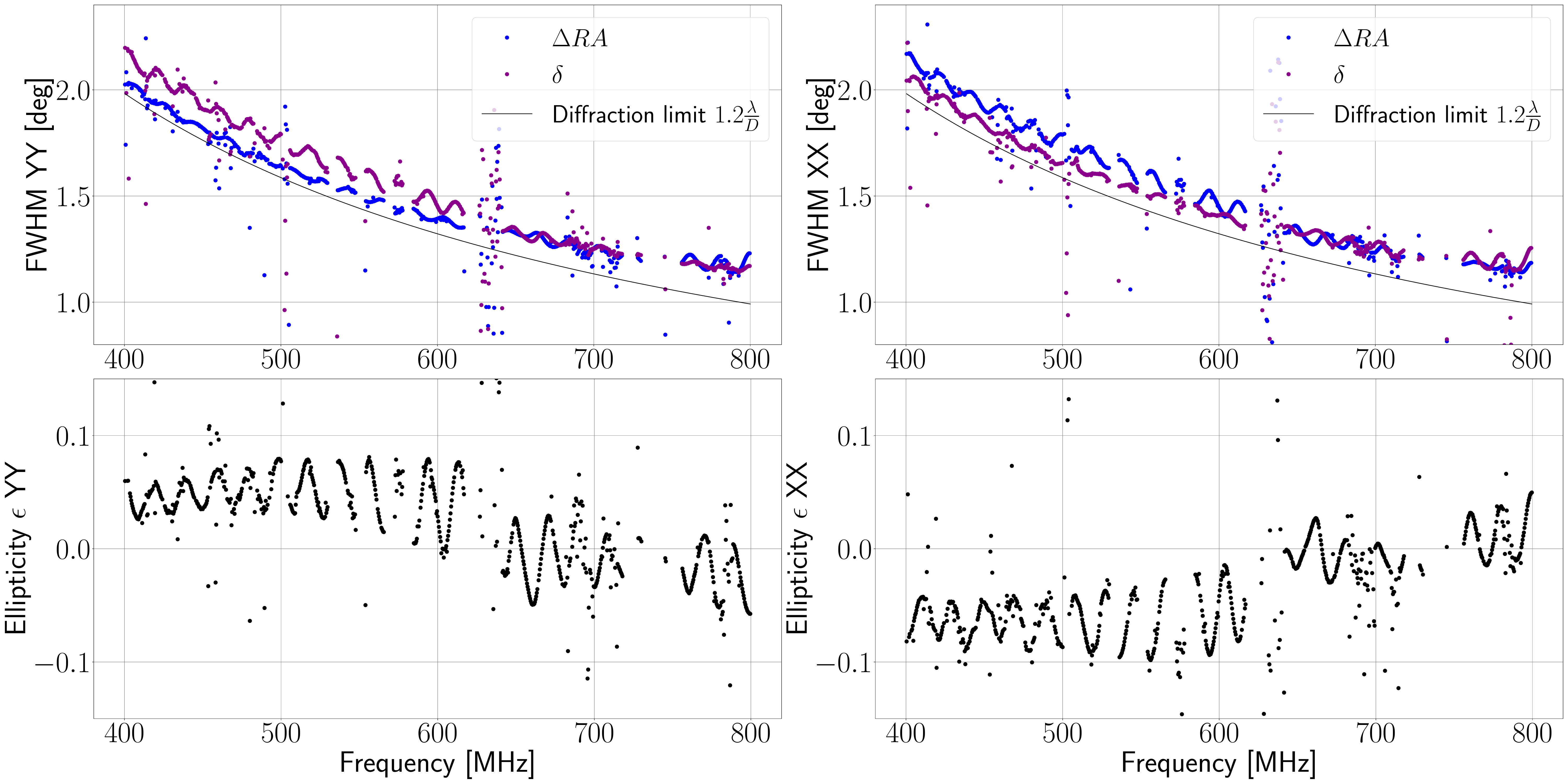}
    \caption{Top row: The best fit beam widths (FWHMs) of the Galt telescope in the East-West (right ascension, blue) and North-South (declination, magenta dots) sky directions as a function of frequency. The black solid curves are the expected FWHM (diffraction limit) of a uniformly illuminated 26\,m diameter aperture as a function of frequency, estimated as $\theta \sim 1.2\frac{\lambda}{D}$. The derived beam widths are systematically worse than diffraction but closely track the trend in frequency before leveling off at high frequencies. The superimposed 20\,MHz ripple is due to standing waves in the cavity between the Galt reflector vertex and the focal structure ($f = 7.6$\,m). The ripples in East-West and North-South are slightly out of phase, resulting in the strong ripple pattern seen in the ellipticities and indicating that this cavity effect is axially asymmetric. Bottom row: The ellipticity, defined as $\epsilon = 2\frac{FWHM_{NS} - FWHM_{EW}}{FWHM_{NS} + FWHM_{EW}}$ of the Galt beam as a function of frequency. At high frequencies the beam is nearly circular, while at lower frequencies there is an average net ellipticity of about 6\%.}
    \label{fig:galtfwhms}
\end{figure*}

The results of this procedure are summarized in Figures \ref{fig:galtbeam717}-\ref{fig:galtfwhms}. Figure \ref{fig:galtbeam717} shows the raw data used for the fit (the colored dots) overplotted on a contour map obtained by evaluating the best fit model, at 717 MHz, in both polarization channels. The beam model and the data have had the best fit background terms removed, and have been normalized by dividing out the best fit amplitude. Figure \ref{fig:galtbeam499} shows the data and model at 499 MHz. Figure \ref{fig:galtfwhms} shows the behavior of the best fit beam-widths (FWHM) in both transverse sky directions as a function of frequency in the top two panels. The bottom two panels show the beam ellipticity as a function of frequency. 

Here we note several of the pertinent features of these results. In the top two panels of Figure \ref{fig:galtfwhms}, the black solid curve is the expected FWHM (diffraction limit) of a uniformly illuminated aperture of 26\,m diameter. The measured beam widths, expected to be worse than diffraction due to the use of a taper, trace the expected trend in frequency, before leveling off at high frequencies. Superimposed on this trend is a ripple pattern with periodicity of about 20\,MHz: this matches what one would expect from a standing wave in the cavity between the vertex of the Galt dish and the feed support structure at a distance of 7.6\,m to the focus. A similar 30 MHz ripple is observed in the FWHM of the CHIME beams (where the CHIME cylinders have $f = 5$\,m). 
We note that the ripples in the East-West and North-South directions are slightly out of phase, indicating there may be slightly different path lengths for radiation entering the cavity in these directions. This latter effect also manifests as a strong ripple pattern in the beam ellipticities, which should otherwise be smooth if the FWHM ripples were in-phase.

Comparing Figures \ref{fig:galtbeam717} and \ref{fig:galtbeam499} it is clear that the beam is noticeably elliptical at lower frequencies. From the lower panels of Figure \ref{fig:galtfwhms}, the average ellipticity in the lower half of the band is $\lesssim 5-6\%$, though it may ripple to as much as $\sim 8-9\%$. This is in excess of what is expected of the beams from  simulations of the custom feed and sleeve structure, which predict the ellipticity to be $\lesssim 2\%$, though previous field measurements with just the feed and sleeve structure 
have 
measured the ellipticity at lower frequencies to be closer to $4\%$ (with considerable uncertainty), such that we anticipate higher ellipticity than the simulations indicated from the feed-sleeve itself. We also note that simulations of the feed-sleeve can not account for additional (azimuthally) asymmetric contributions to the farfield pattern originating from conical-wave scattering off the Galt telescope's three feed support legs \cite{draohigaltbeam}. In any case, uncertainty in the ellipticity does not affect CHIME's outlook for holography, as we observe point sources on boresight and normalize both polarization channels of holography independently. %

\section{Galt system temperature}

\label{section:galtTsys}

\begin{figure*}
    \centering
    \includegraphics[width=1.0\linewidth]{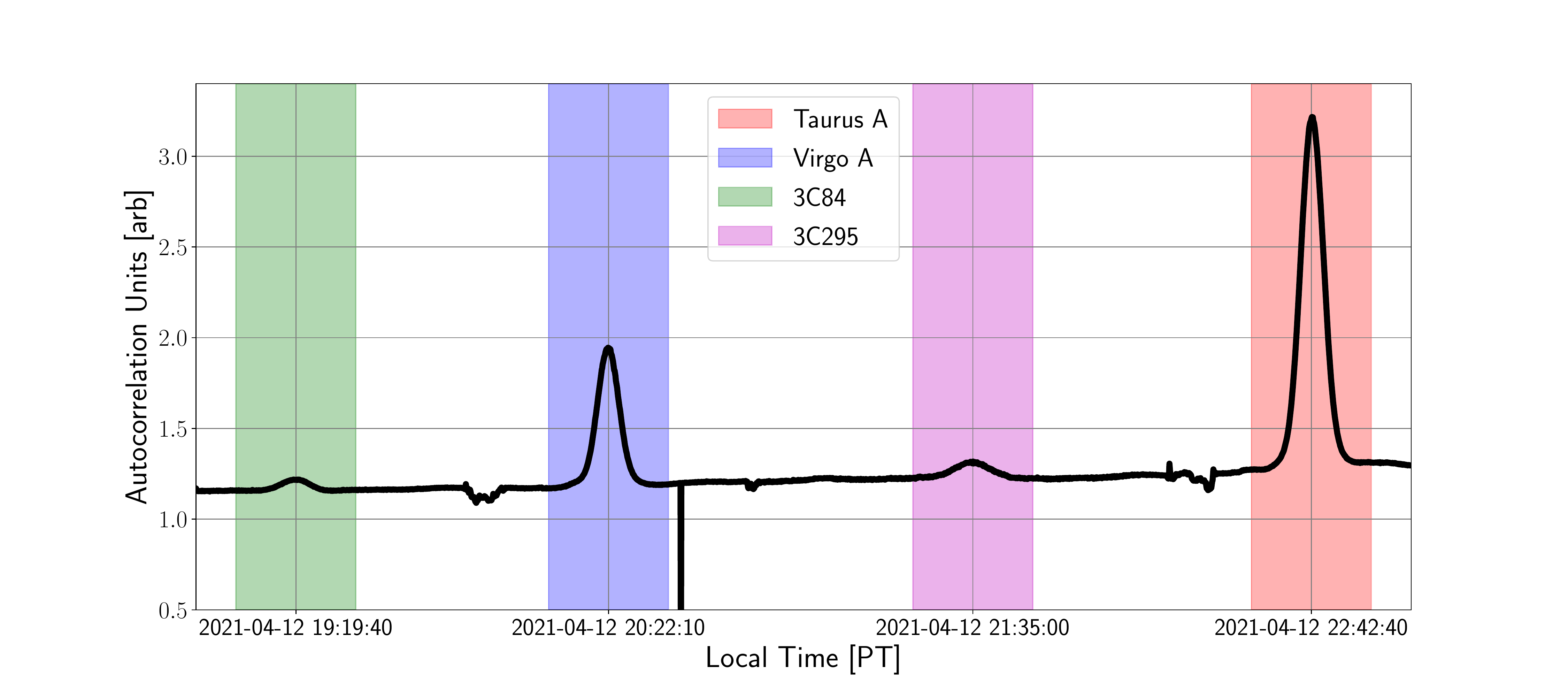}
    \caption{A raw autocorrelation timestream at 717\,MHz in Y polarization of the Galt telescope while performing a driftscan of the radio sources Tau A, Vir A, 3C84, and 3C295. The width of the shaded regions indicates the intervals of data to which we fit the transit profiles to measure the total autocorrelation power.}
    \label{fig:rawdrift}
\end{figure*}

To estimate the system noise of the Galt telescope, a set of four driftscans were performed using the radio sources Taurus A (Tau A), Virgo A (Vir A), 3C84, and 3C295. The Galt telescope was pointed and held in a location ahead of the transit of each source, allowing the source to drift through its beam. Figure \ref{fig:rawdrift} shows an example of the raw driftscan data at 717\,MHz in Y polarization, with the four sources indicated by the shaded regions. 

To estimate the autocorrelation power while a source is transiting through the Galt beam, we fit the transit profiles to a one-dimensional Gaussian, including a flat and linear background contribution. The width of the shaded regions in Figure \ref{fig:rawdrift} indicate the intervals used for fitting. We find that the linear contribution is non-zero but negligible, so for each source we record the autocorrelation power (in arbitrary units) as the sum of the amplitude of the Gaussian profile and the best fit constant background. This is done for all frequencies and both polarization channels. As a check on these results, we compare the derived Gaussian profile widths with the beam-widths (in right ascension) obtained from the Cas A raster scan of Section \ref{section:galtbeam}. The two measurements are found to differ by less than 3\% across the band. 

For our measurement we require estimates of the expected brightness temperatures of the sources plus their backgrounds. The source fluxes are taken from a CHIME-internal catalog of bright radio sources derived from public catalogs provided by the NRAO VLA Sky Survey (NVSS) \cite{nvss} and the VLA Low-frequency Sky Survey (VLSS) \cite{vlss}. These measurements include spectral indices which are used to scale the source fluxes across the entire 400--800\,MHz band. These fluxes are reported in Janskys. To convert to a brightness temperature we require knowledge of the Galt beam shape, for which we use the model derived from our raster scan of Cas A in Section~\ref{section:galtbeam}. %

To estimate the source background temperatures we use the 408 MHz full-sky Haslam map of continuum emission, as reprocessed in Remazeilles et al\cite{haslam}. We examine patches of sky around the four sources used and estimate background by taking the average brightness of the source-subtracted map in a disc around the source location. The width of the disc is set by the geometric mean of the Galt beam-widths in the RA and declination directions 
at 408\,MHz as inferred from the raster scan of Cas A. For all other frequencies we scale the Haslam emission around the four sources by a constant spectral index $\alpha = -0.77$ and use the corresponding Galt beam widths at that frequency. This provides estimates for the radio sky backgrounds of these sources in units of Kelvin. 

To obtain an estimate for the system temperature $T_\text{sys}$ in Kelvin, we apply the Y-factor method \cite{ERAcondonransom}. In this approach, the system under test is connected successively to ``hot" and ``cold" loads of temperatures $T_h$ and $T_c$, respectively, from which one measures the resulting noise powers $P_h$ and $P_c$. The system temperature is then obtained from

\begin{align}
    T_\text{sys} = \frac{T_h - YT_c}{Y - 1} \\
    Y = \frac{P_h}{P_c}
\end{align}

\begin{figure}
    \centering
    \includegraphics[width=1.0\linewidth]{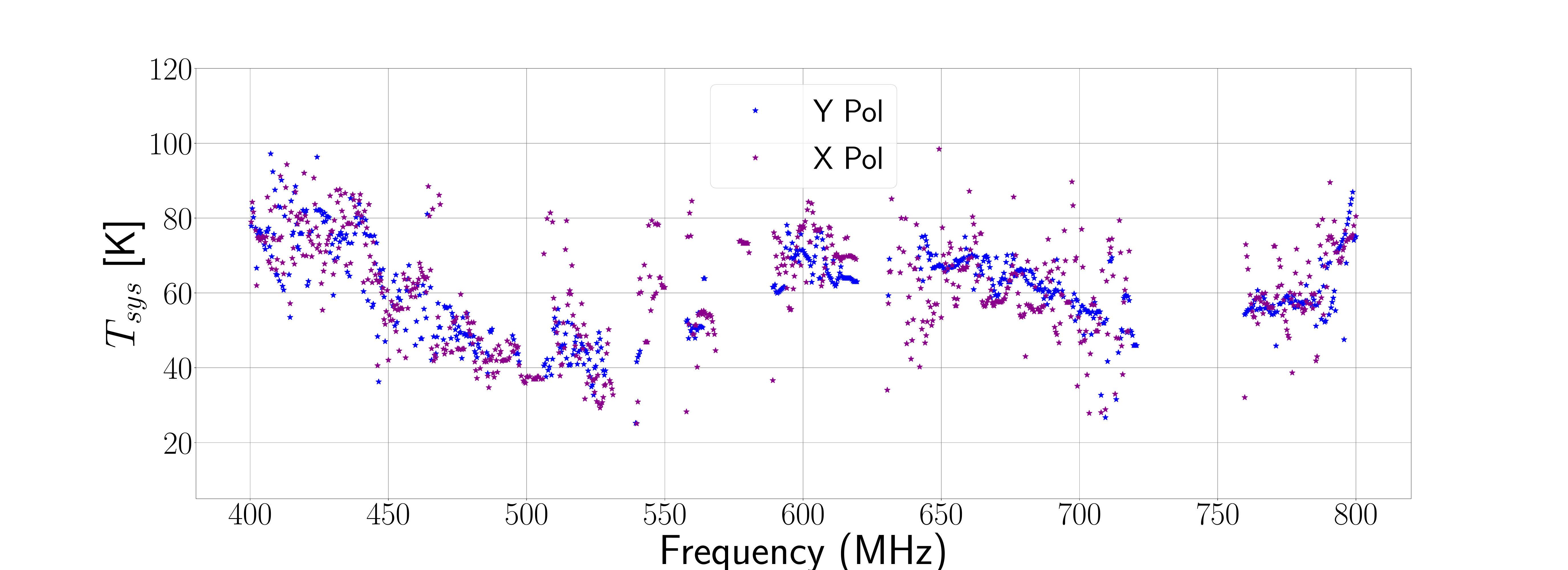}
    \caption{The Galt system temperature, in Y (blue) and X (magenta) polarization as a function of frequency, as measured by the combination of 4 Y-factor measurements using Tau A and Vir A as the hot loads and 3C84 and 3C295 as the cold loads. The measured autocorrelation powers and expected brightness temperatures are estimated as in Section \ref{section:galtTsys}. The most significant overall contribution is from the CHIME LNA immediately following the cloverleaf feed in the Galt receiver chain; the LNA sets the shape of the spectrum, which is not monotonic. Spillover emission from the ground is expected to be a notable secondary contribution.}
    \label{fig:4Yfactors}
\end{figure}

In our case we treat the various celestial sources as the hot and cold loads. We use four different combinations, with Tau A and Vir A as the hot loads and 3C84 and 3C295 as the cold loads. We report the final estimate of the system temperature as the average of these four measurements, and estimate the uncertainty on the measurement as the standard error on the mean ($\lesssim 10$\,K across the band). The results of this procedure are shown in Figure \ref{fig:4Yfactors}. 

On average across the band, the system temperature is about 60--70\,K;
as low as $\sim 40$ K around 500\,MHz, and peaks over 80\,K
at 800\,MHz. As the first element in the Galt receiver chain,
the LNA - whose noise temperature is estimated to be in the
range of 20--30\,K across the band - makes up the most significant overall contribution to the system noise and dominates
the shape of the system noise spectrum. Spillover emission, picked up from the ground off the edges of the reflector, is unaccounted for but expected to be an important secondary contribution. As the Galt reflector has $f / D$ = 0.3, its geometry is expected to incur an additional $\sim -4.5$\,dB of free-space taper in addition to the $-11$\,dB edge taper of the feed cavity design; given a 300\,K ground temperature, spillover emission at the level of $\sim10$\,K would be expected. 

\section{Calibrating Galt Geometry}

\label{section:galtgeo}

\begin{figure*}
    \centering
    \includegraphics[width=1.0\linewidth]{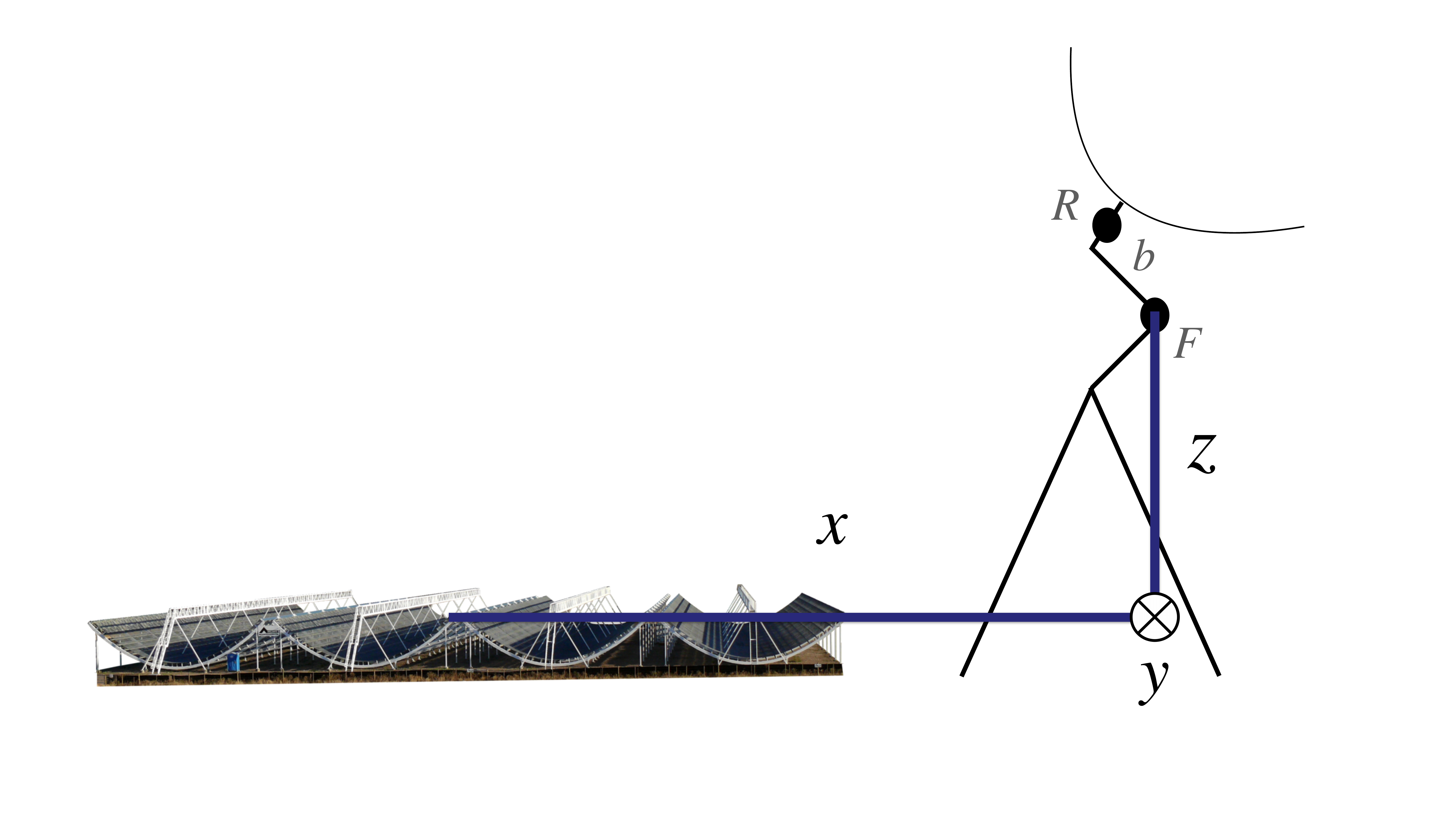}
    \caption{A schematic of a holographic baseline between CHIME (left) and the Galt telescope (shown schematically on the right). CHIME is separated from the fixed point $F$ on the Galt telescope by distances $x, y, z$, measured in meters. The phase center of the Galt telescope for holography, point $R$, is not fixed with pointing; its separation from the fixed point $F$ depends on the observing declination.}
    \label{fig:chimexgaltbsl}
\end{figure*}

\begin{figure*}
    \centering
    \includegraphics[width=1.0\linewidth]{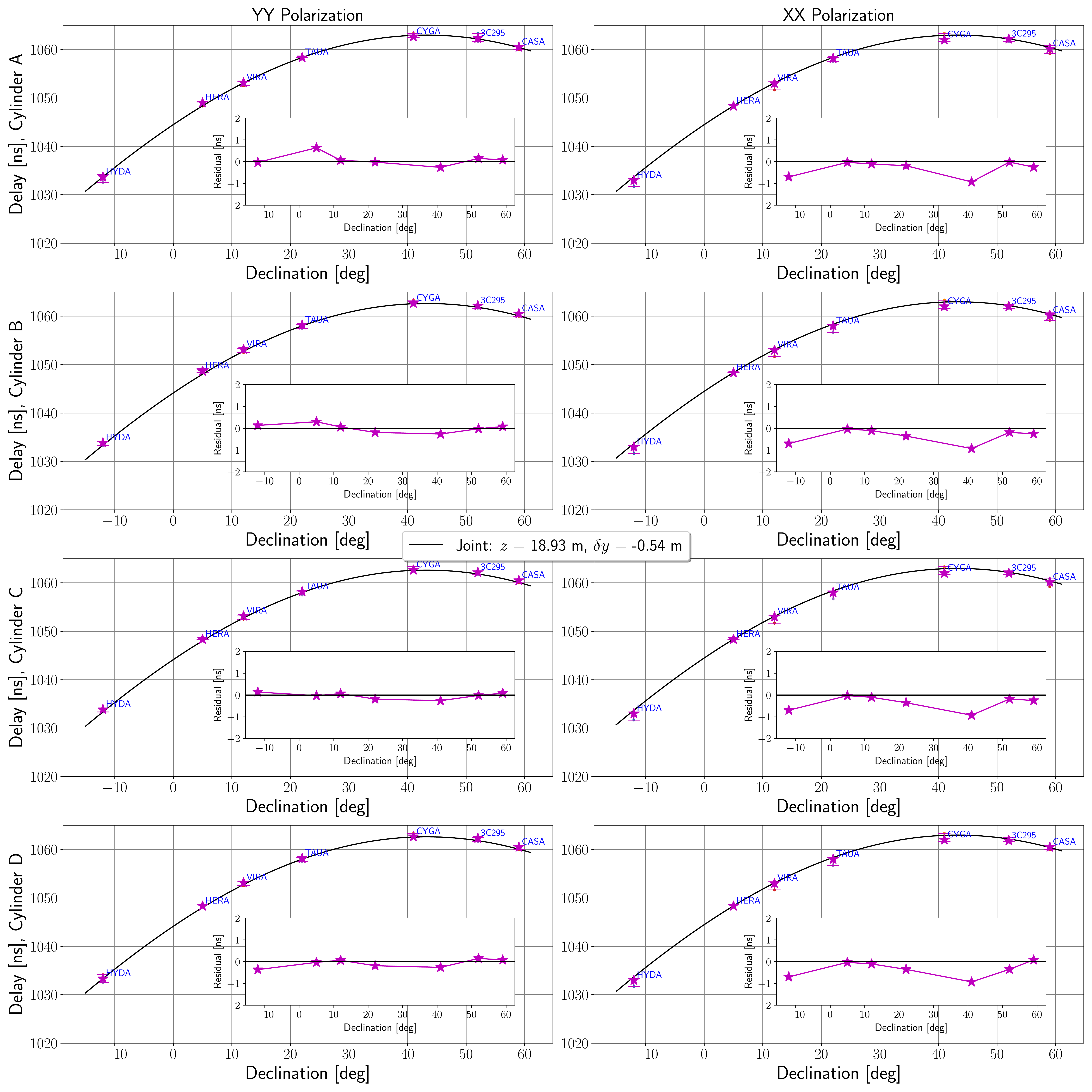}
    \caption{The location of the principal delay peak observed in holography as a function of source declination, averaged over feeds in a cylinder per polarization. We model the data as in Eq \ref{eq:modelfin}, with the result plotted as a solid black curve (for each of the two polarizations, the same curve is fit jointly to all cylinders). The per-cylinder and per-polarization residuals are shown in magenta in inset panels and are below 1 part in $10^3$ across the declination range of this subset of sources.}
    \label{fig:delayresult}
\end{figure*}

To calibrate the interferometric geometry of the Galt dish with CHIME, i.e. to measure the distances separating the Galt phase center and the center of the CHIME array, we perform a delay spectrum analysis of early holography data. 

In interferometry, the physical separation of the two antennas results in the sky signal arriving at the antennas at slightly different times. The relative delay between these times depends on the array layout geometry and the location of the source on the sky. We can therefore learn about the layout of the Galt telescope relative to CHIME by studying the delay structure of the holographic visibilities. This is done using the delay transform, which is implemented as a straightforward Fast Fourier transform of the interferometric time stream from the holographic data set 
along the frequency axis. 

The geometry between CHIME and the Galt telescope is shown schematically in Figure \ref{fig:chimexgaltbsl}, with CHIME on the left and a schematic of the Galt telescope on the right. In the diagram, $F$ and $R$ are, respectively, a fixed point on the telescope (on the right ascension axis of the equatorial mount) and the phase reference point. The fixed component of the holographic baseline is measured between the center of the CHIME array and the fixed point $F$, with East-West, North-South, and vertical separations $x, y, z$. The length $b$ contributes an extra term to the holographic baseline that depends on the declination to which the telescope is pointed. This distance is known from the original design documents of the Galt telescope to be 2.14\,m. 

For this analysis we use holography which has been processed by an early version of our analysis pipeline. This processing consists of a fringestopping step which removes the interferometric phase term, eliminating the rapid fringing of the raw complex timestreams and isolating the beam phase at the declination of the source using an $x$ and $y$ distance previously solved for from the fringe rate. 

Following fringestopping, the data is interpolated onto a common hour angle grid spanning $\pm 60^\circ$ around transit. Early holography was fringestopped without accounting for the vertical displacement $z$ (and the $b$-term) between CHIME and the Galt telescope. We will measure this distance, which we will denote as $z$, as well as residual error in the $y$-separation between CHIME and the Galt telescope. 

Our data selection consists of five observations of each of seven calibrator sources: Hydra A, Hercules A, Vir A, Tau A, Cyg A, 3C295, and Cas A, 35 transits total. This selection of sources spans a wide range of declinations, from $-12^\circ$ (Hydra A) to $+59^\circ$ (Cas A). For each transit, we examine only the time delay on zenith: we can solve for the relevant geometric terms by considering only the zenith, and in addition this maximizes the signal and simplifies the modeling of the geometry. 

We apply several manipulations to the data prior to Fourier transforming. We observe that holography is affected by a feed-to-feed ``jitter" in delay space on the order of 10\,ns. This jitter is corrected for by calibrating the holography with the complex gains derived by the CHIME real-time processing pipeline (see \cite{chimeoverview} for details). We then apply a static mask for radio frequency interference (RFI) which zeroes known bands of persistent contamination from e.g. TV or cellular telephone interference. To improve the resolution of the resulting delay spectrum, we zero-pad the data along the frequency axis by a factor of 4. Finally we use the FFT algorithm as implemented in the python package numpy to calculate the delay transform, then take the complex square to obtain the delay power spectrum. We record the maxima of the delay spectra within a narrow window around the expected location of the peak for all feeds in both polarizations. We median average over feeds common to a cylinder, in each polarization independently. 

In Figure \ref{fig:delayresult} we show the inferred peak delay as a function of source declination. The magenta stars indicate the inferred delay averaged over the 5 observations of that source. We fit this data to the following model for the location of the delay peak as a function of declination:

\begin{align}
    \tau_w(\delta) = \frac{z_{26, CH}}{c}(\sin{LAT}\sin\delta + \cos{LAT}\cos\delta) \\
    \tau_{\Delta v}(\delta) = \frac{\Delta y_{26, CH}}{c}(\cos{LAT}\sin\delta - \sin{LAT}\cos\delta) \\ 
    \tau(\delta) = \tau_w(\delta) + \tau_{\Delta v}(\delta) + \frac{b}{c}\cos\delta + \tau_\text{cable} \label{eq:modelfin}
\end{align}

Above $z_{26, CH}$ is the vertical (towards the sky) distance between the Galt phase center and the center of the CHIME array, $\Delta y_{26, CH}$ is a residual in the relative North-South (in the plane of CHIME) position of the Galt phase center and the center of CHIME, $LAT$ is the latitude of CHIME ($49.32^\circ$), $\tau_\text{cable}$ is the cable delay, and $c$ is the speed of light. $b$ is a term specific to the geometry of the Galt reflector support structure which causes the Galt phase center to drift with declination and must be accounted for to obtain an accurate estimate of $z$ and $\Delta y$. $b$ is known from the original specification of the Galt telescope to be 2.14\,m, so we fix it to this value when performing the fit. We perform the fit by subtracting a model of the form of \ref{eq:modelfin} from all cylinders simultaneously, with a different cable delay parameter for each polarization, then minimizing the joint residuals between both polarizations (that is, we fit for a single value of $z_{26, CH}$, $\Delta y_{26, CH}$, and two cable delays $\tau^{YY}_\text{cable}, \tau^{XX}_\text{cable}$ simultaneously). 

The results of the joint fit are plotted as black curves in Figure \ref{fig:delayresult}, with the model residuals in magenta in the inset panels. The residuals are $\lesssim 1$\,ns. We find a best fit vertical displacement of the Galt dish of 18.93\,m, close to an earlier estimate of 20\,m. We also find a residual in the North South displacement $\delta y$ of -.54\,m. These will be included in revised processing; any error in the geometry of the holographic baselines will appear as a residual in the recovered beam phase. This has potential ramifications for any analysis which depends on the accuracy of the farfield beam phase; for example, analysis of the CHIME reflector structure via the aperture field, obtained by Fourier transform of the holographic visibilities. 

\section*{ACKNOWLEDGMENTS}
We thank Alex Hill, Anna Ordog, and Nikola Milutinovic for their expert operation of the Galt telescope, without which this work would not have been possible. We also extend thanks to Tim Robishaw, Andrew Gray, and the staff of DRAO for coordination of CHIME's observing time on the Galt telescope. We thank Benoit Robert and Rob Messing for their work supporting the operation of the Galt Telescope. We also wish to acknowledge the work of Andre Johnson in designing the custom CHIME receiver for the Galt telescope.  

This material is based on work that has been supported by the NSF through grants (2006548, 2008031, 2006911); by the Perimeter Institute, which is supported in part by the Government of Canada through the Department of Innovation, Science and Economic Development, and by the Province of Ontario through the Ministry of Colleges and Universities; and by the Natural Sciences and Engineering Research Council of Canada, under grant 569654. K.W. Masui holds and acknowledges the support of the Adam J. Burgasser Chair in Astrophysics.

\bibliography{refs}

\begin{thebibliography}{10}

\bibitem{chimeoverview}
{The CHIME Collaboration}, {Amiri}, M., {Bandura}, K., {Boskovic}, A., {Chen},
  T., {Cliche}, J.-F., {Deng}, M., {Denman}, N., {Dobbs}, M., {Fandino}, M.,
  {Foreman}, S., {Halpern}, M., {Hanna}, D., {Hill}, A.~S., {Hinshaw}, G.,
  {H{\"o}fer}, C., {Kania}, J., {Klages}, P., {Landecker}, T.~L., {MacEachern},
  J., {Masui}, K., {Mena-Parra}, J., {Milutinovic}, N., {Mirhosseini}, A.,
  {Newburgh}, L., {Nitsche}, R., {Ordog}, A., {Pen}, U.-L.,
  {Pinsonneault-Marotte}, T., {Polzin}, A., {Reda}, A., {Renard}, A., {Shaw},
  J.~R., {Siegel}, S.~R., {Singh}, S., {Smegal}, R., {Tretyakov}, I., {Van
  Gassen}, K., {Vanderlinde}, K., {Wang}, H., {Wiebe}, D.~V., {Willis}, J.~S.,
  and {Wulf}, D., ``{An Overview of CHIME, the Canadian Hydrogen Intensity
  Mapping Experiment},'' {\em arXiv e-prints} ,  arXiv:2201.07869 (Jan. 2022).

\bibitem{petersontcm}
{Peterson}, J.~B., {Aleksan}, R., {Ansari}, R., {Bandura}, K., {Bond}, D.,
  {Bunton}, J., {Carlson}, K., {Chang}, T.-C., {DeJongh}, F., {Dobbs}, M.,
  {Dodelson}, S., {Darhmaoui}, H., {Gnedin}, N., {Halpern}, M., {Hogan}, C.,
  {Le Goff}, J.-M., {Liu}, T.~T., {Legrouri}, A., {Loeb}, A., {Loudiyi}, K.,
  {Magneville}, C., {Marriner}, J., {McGinnis}, D.~P., {McWilliams}, B.,
  {Moniez}, M., {Palanque-Delabruille}, N., {Pasquinelli}, R.~J., {Pen}, U.-L.,
  {Rich}, J., {Scarpine}, V., {Seo}, H.-J., {Sigurdson}, K., {Seljak}, U.,
  {Stebbins}, A., {Steffen}, J.~H., {Stoughton}, C., {Timbie}, P.~T.,
  {Vallinotto}, A., and {Teche}, C., ``{21-cm Intensity Mapping},'' in [{\em
  astro2010: The Astronomy and Astrophysics Decadal
  Survey}{\nolinebreak\hspace{0.1em}]},   {\bf 2010},  234 (Jan. 2009).

\bibitem{astro-ph/0312134}
{Loeb}, A. and {Zaldarriaga}, M., ``{Measuring the Small-Scale Power Spectrum
  of Cosmic Density Fluctuations through 21cm Tomography Prior to the Epoch of
  Structure Formation},'' {\em \prl}~{\bf 92},  211301 (May 2004).

\bibitem{highztcmroadmap}
{Parsons}, A., {Aguirre}, J.~E., {Beardsley}, A.~P., {Bernardi}, G., {Bowman},
  J.~D., {Bull}, P., {Carilli}, C.~L., {Dai}, W.-M., {DeBoer}, D.~R., {Dillon},
  J.~S., {Ewall-Wice}, A., {Furlanetto}, S.~R., {Gehlot}, B.~K., {Gorthi}, D.,
  {Greig}, B., {Hazelton}, B.~J., {Hewitt}, J.~N., {Jacobs}, D.~C., {Kern},
  N.~S., {Kittiwisit}, P., {Kolopanis}, M., {La Plante}, P., {Liu}, A., {Ma},
  Y.-Z., {Mdlalose}, M., {Mirocha}, J., {Murray}, S.~G., {Nunhokee}, C.~D.,
  {Pober}, J.~C., {Sims}, P.~H., and {Thyagarajan}, N., ``{A Roadmap for
  Astrophysics and Cosmology with High-Redshift 21 cm Intensity Mapping},'' in
  [{\em Bulletin of the American Astronomical
  Society}{\nolinebreak\hspace{0.1em}]},   {\bf 51},  241 (Sept. 2019).

\bibitem{liushawtcmbible}
{Liu}, A. and {Shaw}, J.~R., ``{Data Analysis for Precision 21 cm Cosmology},''
  {\em \pasp}~{\bf 132},  062001 (June 2020).

\bibitem{LEDAintro}
{Price}, D.~C., {Greenhill}, L.~J., {Fialkov}, A., {Bernardi}, G., {Garsden},
  H., {Barsdell}, B.~R., {Kocz}, J., {Anderson}, M.~M., {Bourke}, S.~A.,
  {Craig}, J., {Dexter}, M.~R., {Dowell}, J., {Eastwood}, M.~W., {Eftekhari},
  T., {Ellingson}, S.~W., {Hallinan}, G., {Hartman}, J.~M., {Kimberk}, R.,
  {Lazio}, T. J.~W., {Leiker}, S., {MacMahon}, D., {Monroe}, R., {Schinzel},
  F., {Taylor}, G.~B., {Tong}, E., {Werthimer}, D., and {Woody}, D.~P.,
  ``{Design and characterization of the Large-aperture Experiment to Detect the
  Dark Age (LEDA) radiometer systems},'' {\em \mnras}~{\bf 478},  4193--4213
  (Aug. 2018).

\bibitem{LOFARintro}
{Zaroubi}, S. and {Silk}, J., ``{LOFAR as a probe of the sources of
  cosmological reionization},'' {\em \mnras}~{\bf 360},  L64--L67 (June 2005).

\bibitem{HERAintro}
{DeBoer}, D.~R., {Parsons}, A.~R., {Aguirre}, J.~E., {Alexander}, P., {Ali},
  Z.~S., {Beardsley}, A.~P., {Bernardi}, G., {Bowman}, J.~D., {Bradley}, R.~F.,
  {Carilli}, C.~L., {Cheng}, C., {de Lera Acedo}, E., {Dillon}, J.~S.,
  {Ewall-Wice}, A., {Fadana}, G., {Fagnoni}, N., {Fritz}, R., {Furlanetto},
  S.~R., {Glendenning}, B., {Greig}, B., {Grobbelaar}, J., {Hazelton}, B.~J.,
  {Hewitt}, J.~N., {Hickish}, J., {Jacobs}, D.~C., {Julius}, A., {Kariseb}, M.,
  {Kohn}, S.~A., {Lekalake}, T., {Liu}, A., {Loots}, A., {MacMahon}, D.,
  {Malan}, L., {Malgas}, C., {Maree}, M., {Martinot}, Z., {Mathison}, N.,
  {Matsetela}, E., {Mesinger}, A., {Morales}, M.~F., {Neben}, A.~R., {Patra},
  N., {Pieterse}, S., {Pober}, J.~C., {Razavi-Ghods}, N., {Ringuette}, J.,
  {Robnett}, J., {Rosie}, K., {Sell}, R., {Smith}, C., {Syce}, A., {Tegmark},
  M., {Thyagarajan}, N., {Williams}, P. K.~G., and {Zheng}, H., ``{Hydrogen
  Epoch of Reionization Array (HERA)},'' {\em \pasp}~{\bf 129},  045001 (Apr.
  2017).

\bibitem{HIRAXintro}
{Newburgh}, L.~B., {Bandura}, K., {Bucher}, M.~A., {Chang}, T.~C., {Chiang},
  H.~C., {Cliche}, J.~F., {Dav{\'e}}, R., {Dobbs}, M., {Clarkson}, C., {Ganga},
  K.~M., {Gogo}, T., {Gumba}, A., {Gupta}, N., {Hilton}, M., {Johnstone}, B.,
  {Karastergiou}, A., {Kunz}, M., {Lokhorst}, D., {Maartens}, R., {Macpherson},
  S., {Mdlalose}, M., {Moodley}, K., {Ngwenya}, L., {Parra}, J.~M., {Peterson},
  J., {Recnik}, O., {Saliwanchik}, B., {Santos}, M.~G., {Sievers}, J.~L.,
  {Smirnov}, O., {Stronkhorst}, P., {Taylor}, R., {Vanderlinde}, K., {Van
  Vuuren}, G., {Weltman}, A., and {Witzemann}, A., ``{HIRAX: a probe of dark
  energy and radio transients},'' in [{\em Ground-based and Airborne Telescopes
  VI}{\nolinebreak\hspace{0.1em}]},  {Hall}, H.~J., {Gilmozzi}, R., and
  {Marshall}, H.~K., eds., {\em Society of Photo-Optical Instrumentation
  Engineers (SPIE) Conference Series} {\bf 9906},  99065X (Aug. 2016).

\bibitem{chimedetection}
{CHIME Collaboration}, {Amiri}, M., {Bandura}, K., {Chen}, T., {Deng}, M.,
  {Dobbs}, M., {Fandino}, M., {Foreman}, S., {Halpern}, M., {Hill}, A.~S.,
  {Hinshaw}, G., {H{\"o}fer}, C., {Kania}, J., {Landecker}, T.~L.,
  {MacEachern}, J., {Masui}, K., {Mena-Parra}, J., {Milutinovic}, N.,
  {Mirhosseini}, A., {Newburgh}, L., {Ordog}, A., {Pen}, U.-L.,
  {Pinsonneault-Marotte}, T., {Polzin}, A., {Reda}, A., {Renard}, A., {Shaw},
  J.~R., {Siegel}, S.~R., {Singh}, S., {Vanderlinde}, K., {Wang}, H., {Wiebe},
  D.~V., and {Wulf}, D., ``{Detection of Cosmological 21 cm Emission with the
  Canadian Hydrogen Intensity Mapping Experiment},'' {\em arXiv e-prints} ,
  arXiv:2202.01242 (Feb. 2022).

\bibitem{GBTresult}
{Masui}, K.~W., {Switzer}, E.~R., {Banavar}, N., {Bandura}, K., {Blake}, C.,
  {Calin}, L.~M., {Chang}, T.~C., {Chen}, X., {Li}, Y.~C., {Liao}, Y.~W.,
  {Natarajan}, A., {Pen}, U.~L., {Peterson}, J.~B., {Shaw}, J.~R., and
  {Voytek}, T.~C., ``{Measurement of 21 cm Brightness Fluctuations at z
  \raisebox{-0.5ex}\textasciitilde 0.8 in Cross-correlation},'' {\em
  \apjl}~{\bf 763},  L20 (Jan. 2013).

\bibitem{GBT2022ebossxcorr}
{Wolz}, L., {Pourtsidou}, A., {Masui}, K.~W., {Chang}, T.-C., {Bautista},
  J.~E., {M{\"u}ller}, E.-M., {Avila}, S., {Bacon}, D., {Percival}, W.~J.,
  {Cunnington}, S., {Anderson}, C., {Chen}, X., {Kneib}, J.-P., {Li}, Y.-C.,
  {Liao}, Y.-W., {Pen}, U.-L., {Peterson}, J.~B., {Rossi}, G., {Schneider},
  D.~P., {Yadav}, J., and {Zhao}, G.-B., ``{H I constraints from the
  cross-correlation of eBOSS galaxies and Green Bank Telescope intensity
  maps},'' {\em \mnras}~{\bf 510},  3495--3511 (Mar. 2022).

\bibitem{GMRTresult}
{Chowdhury}, A., {Kanekar}, N., {Das}, B., {Dwarakanath}, K.~S., and {Sethi},
  S., ``{Giant Metrewave Radio Telescope Detection of HI 21 cm Emission from
  Star-forming Galaxies at z = 1.3},'' {\em \apjl}~{\bf 913},  L24 (June 2021).

\bibitem{meerkatresult}
{Cunnington}, S., {Li}, Y., {Santos}, M.~G., {Wang}, J., {Carucci}, I.~P.,
  {Irfan}, M.~O., {Pourtsidou}, A., {Spinelli}, M., {Wolz}, L., {Soares},
  P.~S., {Blake}, C., {Bull}, P., {Engelbrecht}, B., {Fonseca}, J., {Grainge},
  K., and {Ma}, Y.-Z., ``{HI intensity mapping with MeerKAT: power spectrum
  detection in cross-correlation with WiggleZ galaxies},'' {\em arXiv e-prints}
  ,  arXiv:2206.01579 (June 2022).

\bibitem{paperresult}
{Parsons}, A.~R., {Liu}, A., {Aguirre}, J.~E., {Ali}, Z.~S., {Bradley}, R.~F.,
  {Carilli}, C.~L., {DeBoer}, D.~R., {Dexter}, M.~R., {Gugliucci}, N.~E.,
  {Jacobs}, D.~C., {Klima}, P., {MacMahon}, D. H.~E., {Manley}, J.~R., {Moore},
  D.~F., {Pober}, J.~C., {Stefan}, I.~I., and {Walbrugh}, W.~P., ``{New Limits
  on 21 cm Epoch of Reionization from PAPER-32 Consistent with an X-Ray Heated
  Intergalactic Medium at z = 7.7},'' {\em \apj}~{\bf 788},  106 (June 2014).

\bibitem{edgesresult}
{Bowman}, J.~D., {Rogers}, A. E.~E., {Monsalve}, R.~A., {Mozdzen}, T.~J., and
  {Mahesh}, N., ``{An absorption profile centred at 78 megahertz in the
  sky-averaged spectrum},'' {\em \nat}~{\bf 555},  67--70 (Mar. 2018).

\bibitem{saras3result}
{Singh}, S., {Jishnu}, N.~T., {Subrahmanyan}, R., {Udaya Shankar}, N.,
  {Girish}, B.~S., {Raghunathan}, A., {Somashekar}, R., {Srivani}, K.~S., and
  {Sathyanarayana Rao}, M., ``{On the detection of a cosmic dawn signal in the
  radio background},'' {\em Nature Astronomy}~{\bf 6},  607--617 (Feb. 2022).

\bibitem{Seo:2003}
{Seo}, H.-J. and {Eisenstein}, D.~J., ``{Probing Dark Energy with Baryonic
  Acoustic Oscillations from Future Large Galaxy Redshift Surveys},'' {\em
  \apj}~{\bf 598},  720--740 (Dec. 2003).

\bibitem{Seo:2007}
{Seo}, H.-J. and {Eisenstein}, D.~J., ``{Improved Forecasts for the Baryon
  Acoustic Oscillations and Cosmological Distance Scale},'' {\em \apj}~{\bf
  665},  14--24 (Aug. 2007).

\bibitem{1998:Riess}
{Riess, et al.}, A.~G., ``{Observational Evidence from Supernovae for an
  Accelerating Universe and a Cosmological Constant},'' {\em Astronomical
  Journal}~{\bf 116},  1009--1038 (Sept. 1998).

\bibitem{foregrounddayenu}
{Ewall-Wice}, A., {Kern}, N., {Dillon}, J.~S., {Liu}, A., {Parsons}, A.,
  {Singh}, S., {Lanman}, A., {La Plante}, P., {Fagnoni}, N., {Acedo}, E. d.~L.,
  {DeBoer}, D.~R., {Nunhokee}, C., {Bull}, P., {Chang}, T.-C., {Lazio}, T.
  J.~W., {Aguirre}, J., and {Weinberg}, S., ``{DAYENU: a simple filter of
  smooth foregrounds for intensity mapping power spectra},'' {\em \mnras}~{\bf
  500},  5195--5213 (Jan. 2021).

\bibitem{foregroundgprfilter}
{Soares}, P.~S., {Watkinson}, C.~A., {Cunnington}, S., and {Pourtsidou}, A.,
  ``{Gaussian Process Regression for foreground removal in H I Intensity
  Mapping experiments},'' {\em \mnras}~{\bf 510},  5872--5890 (Mar. 2022).

\bibitem{foregroundkpca}
{Irfan}, M.~O. and {Bull}, P., ``{Cleaning foregrounds from single-dish 21 cm
  intensity maps with Kernel principal component analysis},'' {\em \mnras}~{\bf
  508},  3551--3568 (Dec. 2021).

\bibitem{Shawmmodes}
{Shaw}, J.~R., {Sigurdson}, K., {Sitwell}, M., {Stebbins}, A., and {Pen},
  U.-L., ``{Coaxing cosmic 21 cm fluctuations from the polarized sky using m
  -mode analysis},'' {\em \prd}~{\bf 91},  083514 (Apr. 2015).

\bibitem{holographyscottryle}
{Scott}, P.~F. and {Ryle}, M., ``{A rapid method for measuring the figure of a
  radio telescope reflector.},'' {\em \mnras}~{\bf 178},  539--545 (Mar. 1977).

\bibitem{holography30m}
{Morris}, D., {Baars}, J.~W.~M., {Hein}, H., {Steppe}, H., and {Thum}, C.,
  ``{Radio-holographic reflector measurement of the 30-m millimeter radio
  telescope at 22 GHz with a cosmic signal source},'' {\em \aap}~{\bf 203},
  399--406 (Sept. 1988).

\bibitem{holographyeffelsberg100m}
{Godwin}, M.~P., {Schoessow}, E.~P., and {Grahl}, B.~H., ``{Improvement of the
  Effelsberg 100 meter telescope based on holographic reflector surface
  measurement.},'' {\em \aap}~{\bf 167},  390--394 (Oct. 1986).

\bibitem{holographyyebes40m}
López-Pérez, J.~A., de~Vicente~Abad, P., López-Fernández, J.~A.,
  Tercero~Martínez, F., Barcia~Cancio, A., and Galocha~Iragüen, B., ``Surface
  accuracy improvement of the yebes 40 meter radiotelescope using microwave
  holography,'' {\em IEEE Transactions on Antennas and Propagation}~{\bf
  62}(5),  2624--2633 (2014).

\bibitem{1966holotheory}
{Smith}, P., ``{Measurement of the complete far-field pattern of large antennas
  by radio-star sources},'' {\em IEEE Transactions on Antennas and
  Propagation}~{\bf 14},  6--16 (Jan. 1966).

\bibitem{baarsholohistory}
Baars, J. W.~M., ``Metrology of reflector antennas: A historical review,'' {\em
  URSI Radio Science Bulletin}~{\bf 2020}(375),  10--32 (2020).

\bibitem{bergerholo}
{Berger}, P., {Newburgh}, L.~B., {Amiri}, M., {Bandura}, K., {Cliche}, J.-F.,
  {Connor}, L., {Deng}, M., {Denman}, N., {Dobbs}, M., {Fandino}, M.,
  {Gilbert}, A.~J., {Good}, D., {Halpern}, M., {Hanna}, D., {Hincks}, A.~D.,
  {Hinshaw}, G., {H{\"o}fer}, C., {Johnson}, A.~M., {Landecker}, T.~L.,
  {Masui}, K.~W., {Mena Parra}, J., {Oppermann}, N., {Pen}, U.-L., {Peterson},
  J.~B., {Recnik}, A., {Robishaw}, T., {Shaw}, J.~R., {Siegel}, S.,
  {Sigurdson}, K., {Smith}, K., {Storer}, E., {Tretyakov}, I., {Van Gassen},
  K., {Vanderlinde}, K., and {Wiebe}, D., ``{Holographic beam mapping of the
  CHIME pathfinder array},'' in [{\em Ground-based and Airborne Telescopes
  VI}{\nolinebreak\hspace{0.1em}]},  {Hall}, H.~J., {Gilmozzi}, R., and
  {Marshall}, H.~K., eds., {\em Society of Photo-Optical Instrumentation
  Engineers (SPIE) Conference Series} {\bf 9906},  99060D (Aug. 2016).

\bibitem{tomgalt}
{Du}, X., {Landecker}, T.~L., {Robishaw}, T., {Gray}, A.~D., {Douglas}, K.~A.,
  and {Wolleben}, M., ``{Gain and Polarization Properties of a Large Radio
  Telescope from Calculation and Measurement: The John A. Galt Telescope},''
  {\em \pasp}~{\bf 128},  115006 (Nov. 2016).

\bibitem{GaltGMIMS}
{Wolleben}, M., {Landecker}, T.~L., {Douglas}, K.~A., {Gray}, A.~D., {Ordog},
  A., {Dickey}, J.~M., {Hill}, A.~S., {Carretti}, E., {Brown}, J.~C.,
  {Gaensler}, B.~M., {Han}, J.~L., {Haverkorn}, M., {Kothes}, R., {Leahy},
  J.~P., {McClure-Griffiths}, N., {McConnell}, D., {Reich}, W., {Taylor},
  A.~R., {Thomson}, A.~J.~M., and {West}, J.~L., ``{The Global Magneto-ionic
  Medium Survey: A Faraday Depth Survey of the Northern Sky Covering 1280-1750
  MHz},'' {\em \aj}~{\bf 162},  35 (July 2021).

\bibitem{meilingclover}
{Deng}, M. and {Campbell-Wilson}, D., ``{The cloverleaf antenna: A compact
  wide-bandwidth dual-polarization feed for CHIME},'' {\em arXiv e-prints} ,
  arXiv:1708.08521 (Aug. 2017).

\bibitem{CHIMEICE}
{Bandura}, K., {Cliche}, J.~F., {Dobbs}, M.~A., {Gilbert}, A.~J., {Ittah}, D.,
  {Mena Parra}, J., and {Smecher}, G., ``{ICE-Based Custom Full-Mesh Network
  for the CHIME High Bandwidth Radio Astronomy Correlator},'' {\em Journal of
  Astronomical Instrumentation}~{\bf 5},  1641004 (Dec. 2016).

\bibitem{CHIMEGPU}
{Klages}, P., {Bandura}, K., {Denman}, N., {Recnik}, A., {Sievers}, J., and
  {Vanderlinde}, K., ``{GPU Kernels for High-Speed 4-Bit Astrophysical Data
  Processing},'' {\em arXiv e-prints} ,  arXiv:1503.06203 (Mar. 2015).

\bibitem{juanquantizationbias}
{Mena-Parra}, J., {Bandura}, K., {Dobbs}, M.~A., {Shaw}, J.~R., and {Siegel},
  S., ``{Quantization Bias for Digital Correlators},'' {\em Journal of
  Astronomical Instrumentation}~{\bf 7},  1850008--40 (Jan. 2018).

\bibitem{draohigaltbeam}
{Higgs}, L.~A. and {Tapping}, K.~F., ``{The Low-Resolution DRAO Survey of H I
  Emission from the Galactic Plane},'' {\em \aj}~{\bf 120},  2471--2487 (Nov.
  2000).

\bibitem{nvss}
{Condon}, J.~J. e.~a., ``{The NRAO VLA Sky Survey},'' {\em \aj}~{\bf 115},
  1693--1716 (May 1998).

\bibitem{vlss}
{Cohen}, A.~S. e.~a., ``{The VLA Low-frequency Sky Survey (VLSS)},'' in [{\em
  American Astronomical Society Meeting
  Abstracts}{\nolinebreak\hspace{0.1em}]},  {\em American Astronomical Society
  Meeting Abstracts} {\bf 205},  91.05 (Dec. 2004).

\bibitem{haslam}
{Remazeilles}, M., {Dickinson}, C., {Banday}, A.~J., {Bigot-Sazy}, M.~A., and
  {Ghosh}, T., ``{An improved source-subtracted and destriped 408-MHz all-sky
  map},'' {\em \mnras}~{\bf 451},  4311--4327 (Aug. 2015).

\bibitem{ERAcondonransom}
{Condon}, J.~J. and {Ransom}, S.~M.,  [{\em {Essential Radio
  Astronomy}}{\nolinebreak\hspace{0.1em}]} (2016).

\end{thebibliography}
\bibliographystyle{spiebib}

\end{document}